\documentclass[final,5p,times,twocolumn]{elsarticle}

\usepackage{color,balance}

\biboptions{numbers,sort&compress}

\begin{document}

\begin{frontmatter}

\title{How measurement protocols influence the dynamic J-V characteristics of\\ perovskite solar cells: theory and experiment}

\author[1,2]{George Alexandru Nemnes\corref{mycorrespondingauthor}}
\cortext[mycorrespondingauthor]{Corresponding author. Tel.: +40 (0)21 457 4949/157. \\ {\it E-mail address:} nemnes@solid.fizica.unibuc.ro (G.A. Nemnes).}
\address[1]{University of Bucharest, Faculty of Physics, Materials and Devices for Electronics and Optoelectronics Research Center, 077125 Magurele-Ilfov, Romania}
\address[2]{Horia Hulubei National Institute for Physics and Nuclear Engineering, 077126 Magurele-Ilfov, Romania}

\author[3]{Cristina Besleaga}
\address[3]{National Institute of Materials Physics, Magurele 077125, Ilfov, Romania}

\author[3,4]{Andrei Gabriel Tomulescu}
\address[4]{University of Bucharest, Faculty of Physics, 077125 Magurele-Ilfov, Romania}

\author[3]{Alexandra Palici}

\author[3]{\\Lucian Pintilie}

\author[5]{Andrei Manolescu}
\address[5]{School of Science and Engineering,
        Reykjavik University, Menntavegur 1, IS-101 Reykjavik, Iceland}

\author[3]{Ioana Pintilie\corref{mycorrespondingauthor2}}
\cortext[mycorrespondingauthor2]{Corresponding author. Tel.: +40 (0)21 2418 230. \\ {\it E-mail address:} ioana@infim.ro (Ioana Pintilie).}

\begin{abstract}
The dynamic effects observed in the J-V measurements represent one important hallmark in the behavior of the perovskite solar cells. Proper measurement protocols (MPs) should be employed for the experimental data reproducibility, in particular for a reliable evaluation of the power conversion efficiency (PCE), as well as for a meaningful characterization of the type and magnitude of the hysteresis. We discuss here several MPs by comparing the experimental J-V characteristics with simulated ones using the dynamic electrical model (DEM). Pre-poling conditions and bias scan rate can have a dramatic influence not only on the apparent solar cell performance, but also on the hysteretic phenomena. Under certain measurement conditions, a hysteresis-free behavior with relatively high PCEs may be observed, although the J-V characteristics may be far away from the stationary case. Furthermore, forward-reverse and reverse-forward bias scans show qualitatively different behaviors regarding the type of the hysteresis, normal and inverted, depending on the bias pre-poling. We emphasize here that correlated double-scans, forward-reverse or reverse-forward, where the second scan is conducted in the opposite sweep direction and begins immediately after the first scan is complete, are essential for a correct assessment of the dynamic hysteresis. In this context, we define a hysteresis index which consistently assigns the hysteresis type and magnitude. Our DEM simulations, supported by experimental data, provide further guidance for an efficient and accurate determination of the stationary J-V characteristics, showing that the type and magnitude of the dynamic hysteresis may be affected by unintentional pre-conditioning in typical experiments.
\end{abstract}


\end{frontmatter}


\section{Introduction}

In the past few years the perovskite solar cells (PSCs) witnessed an impressive development in terms of reported power conversion efficiencies (PCEs) \cite{PIP:PIP2855}. In spite of the rapid advancements, early studies already pointed out the rather unusual hysteretic behavior \cite{snaith}, which creates difficulties for a correct PCE determination. The J-V hysteresis is a dynamic effect typically influenced by the pre-poling conditions, bias scan rate, scan direction and measurement history \cite{doi:10.1021/acs.jpclett.5b00289,doi:10.1021/acs.jpclett.6b00215,doi:10.1021/jacs.7b10958}. More recently, an increasing number of studies indicate enhanced PCEs in connection with a diminished or even hysteresis-free behavior. This is typically achieved by improving the electron extraction and by reducing the number of surface traps, employing fullerene derivatives \cite{Xu2015,C5EE00120J,doi:10.1021/acs.jpclett.6b02103}, electron selective layers such as SnO$_2$ \cite{AENM:AENM201700414,C7TA08040A,doi:10.1002/adfm.201706276} or ZnO \cite{doi:10.1002/adma.201705596}, by optimizing the growth of the perovskite active layer \cite{doi:10.1021/acsenergylett.8b00871,SIDHIK2017169}, by impurity co-doping of the electron transfer layer \cite{doi:10.1021/acsami.7b16312} or of the perovskite absorber \cite{C7EE02901B}.
However, in general, as the employed measurement procedures are rather different and, in some cases, insufficiently controlled or specified, it is quite difficult to assess and compare the potential hysteretic effects: to what extent the hysteretic effects or their absence are truly related to fabrication methods or are a consequence of the measurement conditions? Since the dynamic hysteresis may be further linked to solar cell degradation \cite{Calado2016,C6TA09202K,doi:10.1021/acs.jpclett.6b02375,Aristidou2017}, a detailed knowledge of the solar cell operation and the underlying mechanisms is crucial in the development of commercially ready PCSs.  
In this context, standardized measurement protocols (MPs) should be established in order to reliably extract the photovoltaic performance indicators as well as to select the most promising candidates with potential for enhanced stability. 

A set of guidelines for proper characterization of PSCs were outlined by {\it Christians et al.}, based on recurring dynamic hysteretic effects in the J-V characteristics, e.g. the impact of the scan rate and poling voltage bias \cite{doi:10.1021/acs.jpclett.5b00289}. {\it Zimmermann et al.} proposed a MP in five steps, which includes a standard J-V measurement with forward and reverse individual scans, steady state tracking, cyclic and time-resolved J-V measurements followed by a repeated first step \cite{doi:10.1063/1.4960759}. 
More recently, a detailed discussion regarding reliable measurement techniques was provided by {\it Dunbar et al.}, presenting a broad inter-laboratory comparison \cite{C7TA05609E}. Here, the authors propose a strategy for PCE determination by identifying the most appropriate technique, depending on the magnitude of the degradation and stabilization time scales: maximum-power-point tracking (MPPT), stabilized current at fixed voltage (SCFV) or dynamic I-V measurements. Further on, {\it Pellet} et al. explore the difficulties posed by standard MPPT algorithms such as perturb and observe \cite{PIP:PIP2894} due to the slow response of the PSCs and argue for further investigations in PCE tracking. 
\begin{figure*}[t]
\centering
\includegraphics[width=16.5cm]{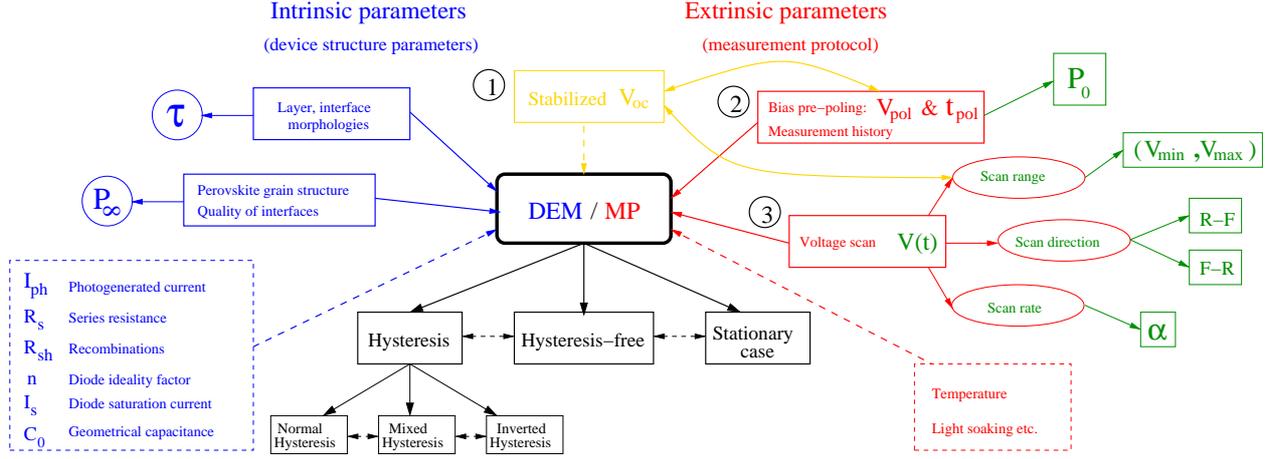}
\caption{Schematics of the influence of the intrinsic and extrinsic parameters on the dynamic hysteretic phenomena as described by DEM, within a specified MP. The experimental parameters in focus here are the pre-poling voltage bias $V_{\rm pol}$ applied for a time $t_{\rm pol}$ (i.e. initial polarization $P_0$ in DEM) and the parameters related to the voltage scan $V(t)$ (scan rate, direction and range). The actual poling of the PSC at fixed scan rate and direction depends on $V_{\rm pol}$ and scan range $(V_{\rm min},V_{\rm max})$ in connection to the stabilized open circuit bias $V_{\rm oc}$. Several hysteresis types may be observed: normal hysteresis (NH), inverted hysteresis (IH) or mixed hysteresis (MH). The dynamic effects are also strongly influenced by the structure dependent intrinsic parameters, particularly by the relaxation time $\tau$ and open circuit polarization $P_\infty$. Our three-step MP requires that the scan range and rate and, optionally, the poling voltage are correctly set, in relation to the pre-determined $V_{\rm oc}$. Color codes: model parameters (blue), experimental parameters (red), $V_{\rm oc}$ stabilization step (orange), MP specific parameters which are also found in DEM (green) and the dynamic J-V outcome (black).}
\label{diagram}
\end{figure*}

The dynamic effects introduce inherent difficulties in the reproducibility of the experimental data, as the initial conditions 
are essential, but quite hard to control, especially when specific time constants are comparable with the measurement time interval or the time spent between measurements. It was recently established that normal hysteresis (NH) can be systematically turned into inverted hysteresis (IH) by switching the initial polarization, which is achieved by changing the pre-poling bias from above the open circuit bias to negative values \cite{doi:10.1021/acs.jpcc.7b04248}. NH and IH \cite{AENM:AENM201600396} correspond to a counter-clock-wise and clock-wise evolutions of the J-V characteristics in the first quadrant, respectively. A unified picture of these two apparently independent hysteretic phenomena was provided by the dynamic electrical model (DEM) \cite{NEMNES2017197}, which introduces a time dependent description of the current and internal polarization, as the PSC is subjected to an arbitrary voltage scan and initial poling conditions. DEM renders the time evolution of the polarization charge, which strives for a bias dependent equilibrium value, within a relaxation time of the order of seconds. 
A similar approach was formulated for the surface polarization voltage, where it was hypothesized that cations accompanied by carrier accumulations are responsible for the hysteretic effects \cite{doi:10.1021/acs.jpclett.7b00045}.
Presently, there is a rather broad agreement that the slow process is most likely linked to ion migration \cite{Eames2015,doi:10.1021/acs.jpclett.7b00975}, although no direct experimental proof of process reversibility was reported so far. 

In this paper we address some key aspects in the MPs that have not been systematically discussed so far, but are essential for an accurate description of the hysteretic phenomena. Although the importance of solar cell pre-conditioning is widely recognized, independent forward and reverse scans are typically performed within a fixed bias range, without a definite control over the initial poling conditions for each scan direction. The time between measurements is another issue that needs careful consideration as well as the influence of the measurement history. In this context, we propose a class of MPs based on correlated forward and reverse bias scans, which enables a consistent evaluation of the dynamic hysteresis and, in particular, an efficient and accurate measurement of the stationary J-V characteristics. Here, correlated J-V scans are double-scans where the second scan begins immediately after the first scan is complete. Importantly, forward-reverse (F-R) and reverse-forward (R-F) scans are shown to be non-equivalent with respect to the type of the hysteresis, normal or inverted, depending on the PSC pre-poling. We characterize the dynamic hysteresis by introducing a new hysteretic index. Other factors like a too large or too small bias scan rate can render a probe hysteresis-free, although it may display large hysteresis for intermediate scan rates. However, all the apparently conflicting features observed in different types of measurements are suitably explained by the DEM model. Thus, by connecting the dynamic J-V measurements to DEM simulations we obtain a comprehensive overview regarding the dynamic regime and an accurate prediction over the experimental results. Going beyond the stationary regime, as a routine procedure, may provide additional information about the structure and evolution of the PSC, e.g. an early assessment regarding potential degradation.

\section{Discussion}

\subsection{MPs from DEM perspective}

Based on DEM we may describe the actual outcome of a measurement conducted within specified experimental conditions. 
In order to have an overview of the subsequent discussion, we present in Fig.\ \ref{diagram} a diagram describing the relations between DEM together with the employed MP and the result of the measurement, having as input experimentally accessible parameters. Depending on the MP and the physical parameters, the hysteretic behavior can change from normal hysteresis to inverted hysteresis, with an intermediate mixed hysteresis (MH) behavior, characterized by one or more crossing points between the forward and the reverse characteristics in $(0,V_{\rm oc})$ range. Similarly, under certain measurement or sample preparation conditions, a hysteresis-free behavior may be observed, which may not always coincide with the stationary characteristics. In this context, as indicated in Fig.\ \ref{diagram}, we divide the set of input parameters in two categories. The first one corresponds to the structure related parameters, called intrinsic parameters, such as the relaxation time scale $\tau$, the polarization at open-circuit bias $P_\infty$, the photo-generated current under standard AM 1.5 illumination conditions $I_{\rm ph}$, series and shunt resistances, $R_{\rm s}$ and $R_{\rm sh}$, diode ideality factor $n$, diode saturation current $I_{\rm s}$ and geometrical capacitance $C_0$. The second group of parameters, i.e. the extrinsic parameters, controls the measurement conditions: the initial polarization $P_0$ set by poling the PSC at $V_{\rm pol}$ for a time $t_{\rm pol}$, the voltage scan $V(t)$, within the scan range $(V_{\rm min},V_{\rm max})$, scan rate $\alpha=dV/dt$, and scan direction, i.e. R-F or F-R scans, and other external parameters and pre-conditioning steps like temperature, light soaking etc.

It is already known that the dynamic J-V characteristics in either forward or reverse bias scans depend strongly on the bias scan rate $\alpha$. Tress et al. \cite{C4EE03664F} showed that in the R-F scans the short-circuit current is enhanced by increasing $\alpha$, while the hysteresis is minimized for very large and for small scan rates. In addition, the current overshoot in the reverse characteristics was observed in other studies \cite{meloni2015,C5EE02740C} as well. These features were systematically explained by DEM and confirmed by experiment \cite{NEMNES2017197}, while further studies substantiate this picture \cite{doi:10.1021/acs.jpclett.7b00045}. Our theoretical framework is mapped onto a three step MP introduced in Ref.\ \cite{doi:10.1021/acs.jpcc.7b04248}, which includes $V_{\rm oc}$ stabilization, pre-poling phase, i.e. applying $V_{\rm pol}$ for a time interval $t_{\rm pol}$, and actual measurement performed in $(0,V_{\rm oc})$ bias range. In addition to the R-F scans we investigate here also the F-R scans and establish a unified picture regarding the type and magnitude of the hysteresis, depending on both poling and scan direction. The following analysis of the dynamic effects, from both theoretical and experimental point of view, has a two-fold outcome: a correct determination of the maximum PCE by performing voltage bias scans starting from the corresponding stationary poling conditions and a consistent evaluation of the dynamic J-V hysteresis using proper MPs.

\subsection{DEM simulations}

In order to distinguish between genuine dynamic effects and potential degradation, which may be both present in a real PSC, we start our analysis by considering an ideal cell and explore in the framework of DEM simulations the effects induced by pre-conditioning in the dynamic J-V characteristics. The DEM is based on the coupled system of differential equations \cite{NEMNES2017197,doi:10.1021/acs.jpcc.7b04248}, numerically solved using {\it GNU Scientific Library} (GSL) \cite{gsl}:
\begin{eqnarray}
\label{Idiffeq2}
-R_{\rm s} C_0 \frac{\partial I}{\partial t} &=&
      I_{\rm s} \left( e^{\frac{q(V + I R_{\rm s})}{n k_{\rm B} T}}-1\right) 
+ \left(\frac{R_{\rm s}}{R_{\rm sh}} + 1 \right) I \nonumber \\
& &+ \frac{V}{R_{\rm sh}} + C_0 \frac{\partial V}{\partial t}
    + {\mathcal A}\frac{\partial P_{\rm nl}}{\partial t} - I_{\rm ph}\ , 
\end{eqnarray}
\begin{equation}
\frac{\partial P_{\rm nl}}{\partial t} = - \frac{P_{\rm nl}(t) - P_{\rm nl,\infty}(U_{\rm c}(t))}{\tau}\ ,
\label{dPdt}
\end{equation}
which includes as parameters the series resistance $R_{\rm s}$, the shunt resistance $R_{\rm sh}$, the diode saturation current $I_{\rm s}$ and diode ideality factor $n$, the photogenerated current $I_{\rm ph}$, the geometrical capacitance $C_0$ and the device area ${\mathcal A}$. The evolution of the time-depending quantities, namely the current $I$ and the non-linear polarization $P_{\rm nl}$, are determined by the employed bias scan procedure and the initial poling conditions, $P_0 = P_{\rm nl}(t=0)$. The stationary value of the non-linear polarization $P_{\rm nl,\infty}= (V + I R_{\rm s})/V_{\rm oc} P_{\rm \infty}$ depends on the applied voltage on the capacitor $U_{\rm c} = V + I R_{\rm s}$, where $V_{\rm oc}$ is the open-circuit bias and $P_{\rm \infty}$ represents the stationary polarization at open-circuit.

\begin{figure}[t]
\centering
\includegraphics[width=8.5cm]{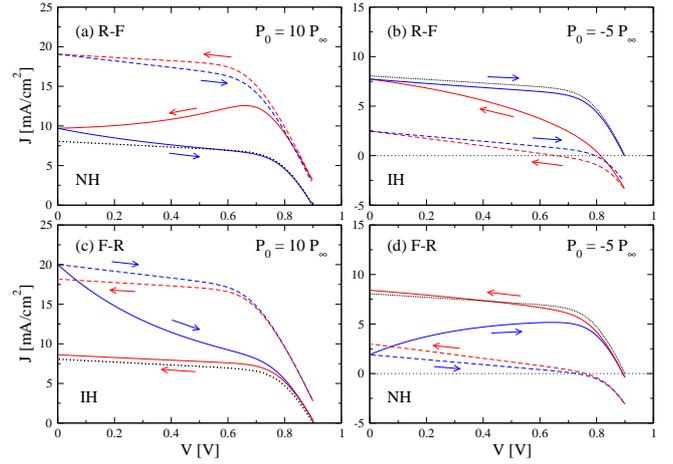}
\caption{Simulated J-V characteristics for continuous forward and reverse bias scans: R-F scans (a,b) and F-R scans (c,d) for two bias scan rates, $\alpha = 20$ mV/s (solid) and $500$ mV/s (dashed), and two poling conditions, $P_0 = 10 P_\infty$ and $-5 P_\infty$. The reverse and forward scans are depicted in red and blue color, respectively. NH (a,d) and IH (b,c) behavior depends both on the sign of $P_0$ and on the scan direction, R-F or F-R. The stationary J-V characteristics is plotted with dotted lines.}
\label{RF-FR}
\end{figure}

Figure\ \ref{RF-FR} shows R-F and F-R scans performed with two different scan rates, $\alpha =$ 20 mV/s and 500 mV/s, while the PSCs are pre-conditioned with both positive and negative initial polarizations, $P_0 = 10 P_\infty$ and $P_0 = -5 P_\infty$. For the simulated J-V characteristics we used the following parameters, calibrated on a typical experimental PSC device: $J_{\rm ph}=8.22$ mA/cm$^2$, $R_{\rm s} = 120$ $\Omega$, $R_{\rm sh} = 6$ k$\Omega$, $n = 1.55$, $J_{\rm s} =1.1$ pA/cm$^2$, $P_\infty = 25$ mC/cm$^2$, $\tau=20$ s and a negligible geometrical capacitance $C_0$. We consider $k_{\rm B} T = 26$ meV. The R-F scan with a large positive pre-poling shown in Fig.\ \ref{RF-FR}(a) presents significant NH for the rather slow scan of 20 mV/s. By contrast, the hysteresis is significantly diminished for the fast scan of 500 mV/s. However, the J-V characteristics are far away from the stationary case and, regarding the fast scan, this would inaccurately correspond to a PSC with relatively high PCE and small hysteresis.
Besides NH we showed recently that an inverted hysteresis \cite{doi:10.1021/acs.jpcc.7b04248} may occur when the initial polarization is negative. Unlike to the previous case, as depicted in Fig.\ \ref{RF-FR}(b), the reverse and also the short-circuit current ($I_{\rm sc}$) are reduced, which would lead to an underestimation of the PCE. Moreover, $I_{\rm sc}$ becomes smaller while increasing $\alpha$, while the same trend of diminished hysteresis at very fast scans is obtained as for positive poling. 

Next we investigate in how far changing the order of forward and reverse bias scans is important for the type and magnitude of the hysteresis. To this end we consider F-R scans with the same initial polarizations, depicted in Fig.\ \ref{RF-FR}(c,d). Under these conditions, in contrast to R-F scan, IH appears for $P_0>0$ and NH is found for $P_0<0$, as the initial polarization (absolute value) decays mainly on the forward scan. Here, a comment should be made regarding the relaxation time $\tau$ in relation to the measurement interval, as given by $\alpha$, for the F-R scan with $P_0>0$: as long as $\tau$ is large enough, IH behavior is obtained; however if $\tau$ is small, MH may appear as the initial polarization decays fast and the NH is recovered at higher biases. One should also note that for all four measurement conditions employing consecutive F-R and R-F scans, in the case of a slow scan rate ($\alpha = 20$ mV/s), always the second bias scan, either forward or reverse, is closest to the stationary case, as the initial pre-poling effect is neutralized to a large extent during the first scan. Here, by the choice of the simulations parameters, completely inverted hysteresis for both R-F ($P_0 = -5 P_\infty$) and F-R ($P_0 = 10 P_\infty$) scans is obtained.

\begin{figure}[t]
\centering
\includegraphics[width=8.5cm]{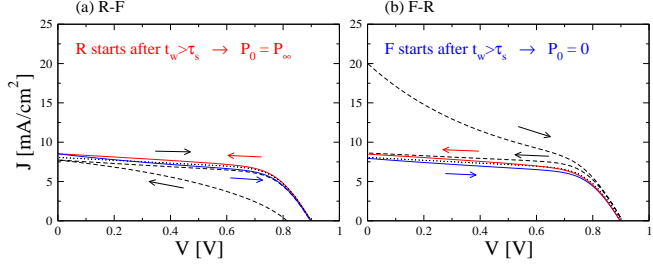}
\caption{J-V characteristics corresponding to stationary poling conditions: (a) $P_0=P_\infty$ for R-F and (b) $P_0=0$ for F-R, at a scan rate of $20$ mV/s. This may be achieved while waiting enough time, $t_{\rm w}>\tau_{\rm s}$, at $V=V_{\rm oc}$ (R-F) and $V=0$ (F-R), after accidental sample poling. The dashed lines, corresponding to the data sets in Fig.\ \ref{RF-FR} (b) and (c), respectively, for the same bias scan rate, were added for comparison, while the stationary J-V characteristics is represented by dotted lines. The type and magnitude of the hysteresis may change according to the time interval $t_{\rm w}$.}
\label{tw}
\end{figure}

Analyzing the type and magnitude of the dynamic hysteresis implies a direct comparison between forward and reverse scans. Correlated forward and reverse bias scans, which have a relation between the two values of the initial polarizations, are essential for a consistent description. On the other hand, uncorrelated forward and reverse bias scans may not provide an accurate characterization of the dynamic hysteresis, since controlling the initial poling conditions of each individual scan is rather difficult. Conversely, consecutive R-F or F-R scans exclude potential variations in polarization/depolarization of the PSC in the time interval spent between the two scans. 
In many instances, a MH behavior with one crossing point at $V_{c}$ is found, which depends on several parameters, extrinsic and intrinsic, like the pre-poling $P_0$, scan rate $\alpha$ and relaxation time scale $\tau$, as detailed in Fig. S1 the Supplementary Material (SM). 

A correct assessment of the stationary J-V characteristics is essential for an accurate and efficient determination of the maximum PCE and, in this context, the pre-poling of the PSC is undesired. Although MPPT or SCFV techniques can be locally more reliable, getting the optimal J-V parameters often requires the exploration of a finite bias range in advance, which is performed by dynamic J-V scans. Typical measurements use a fixed bias range $(V_{\rm min},V_{\rm max})$, with $V_{\rm min}<0$ and $V>V_{\rm oc}$ in order to capture the entire range of working conditions. However, using a fixed bias scan range, without relating it to $V_{\rm oc}$, which may slightly change over time, will inadvertently induce a certain degree of bias pre-poling, i.e. corresponding to the time interval for which the measurement takes place outside $(0,V_{\rm oc})$ bias range. To obtain a J-V characteristics as close as possible to the stationary case, we consider the stationary poling conditions are achieved, $P_0=P_\infty$ for R-F and $P_0=0$ for F-R, as indicated in Fig.\ \ref{tw}. The scan rate should be as low as possible, but without interfering with the characteristic time interval of the potential degradation process \cite{doi:10.1021/acs.jpclett.6b02375}.
Taking into account a possible pre-poling from a previous scan, the depolarization of the PSC can be pursued by stabilizing the cell at $V=V_{\rm oc}$ (R-F) or at $V=0$ (F-R). This may be achieved by keeping the bias for a waiting time between measurements, $t_{\rm w}$, larger than stabilization time, $\tau_{\rm s}$. In this way, the PSC polarization is reinitialized to a large degree and one obtains minimal hysteresis J-V characteristics close to the stationary case. 
Both scans corresponding to stationary poling conditions present a rather small NH, its magnitude being set by $P_\infty$ and the finite $\alpha$.  
In the R-F scan, negative initial poling can have as outcome a relatively large IH for small $t_{\rm w}$, while for long enough waiting times it can change to a small NH. A similar situation is found for the F-R scan with $P_0>0$. Therefore, the type of the hysteresis can change simply by adjusting the time between measurements, which may have had introduced undesired additional polarization. 
By contrast, when considering R-F scans with accidental positive poling or F-R scans with negative poling, only the magnitude of NH hysteresis may change. 
The sensitivity of the magnitude and type of hysteresis to $t_{\rm w}$ in relation to $\tau_{\rm s}$, poling conditions and scan direction should be therefore taken into account in a consistent manner in order to achieve meaningful and reproducible results.       
The tunability of the hysteresis was also reported in other studies, but this was connected to the structural configuration of the PSC, such as the stoichiometry of the perovskite layer \cite{doi:10.1021/acs.jpclett.7b00571} or modification of the c-TiO$_2$ layer \cite{C7EE02048A}. It remains an open question to what extent the hysteretic behavior of PSCs exhibiting structure dependent hysteresis can be further tuned by bias pre-poling and measurement conditions.

\subsection{Experimental validation}

Following the DEM simulations discussed in Fig.\ \ref{RF-FR}, we reproduce the corresponding experimental conditions for the R-F and F-R scans. The fabrication of the PSCs and the characterization methods are described in the SM.

\begin{figure}[t]
\centering
\includegraphics[width=8.5cm]{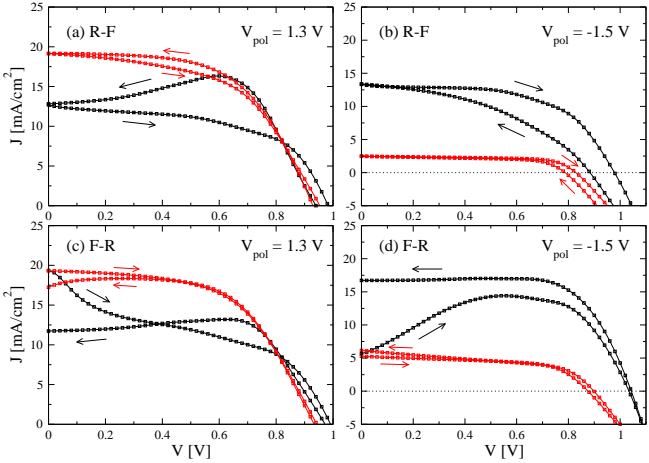}
\caption{Experimental J-V characteristics obtained by R-F (a,b) and F-R (c,d) scans, at the scan rates $\alpha=$ 20 mV/s (black) and 500 mV/s (red). Pre-poling was performed at $V_{\rm pol}=$ 1.3V and -1.5V, for a time $t_{\rm pol}=$ 30s.}
\label{RF-FR-exp}
\end{figure}

The J-V measurements shown in Fig.\ \ref{RF-FR-exp} are performed at the same bias scan rates, $\alpha=$ 20 mV/s and 500 mV/s, while the initial polarization $P_0$ is achieved by pre-conditioning the solar cell at $V_{\rm pol}$ for a time $t_{\rm pol}$, employing the three step MP introduced in Ref.\ \cite{doi:10.1021/acs.jpcc.7b04248}. We apply $V_{\rm pol}=$ 1.3V and -1.5V for $t_{\rm pol}=30$ s yielding positive and negative polarizations, respectively. One should mention that, alternatively, the bias pre-poling may occur, e.g. in an R-F scan, simply by taking the start bias ($V_{\rm max}$) significantly larger than $V_{\rm oc}$, and performing a reverse scan with constant rate \cite{NEMNES2017197}, which is a more typical situation found in usual measurements. The experimental J-V characteristics have the same features as the simulated ones: an overall increase of the current for positive poling in contrast to negative poling and, in the limit of large scan rates, the hysteresis is diminished. The hysteresis types match the ones in the DEM simulations: NH is found for $V_{\rm pol}>0$ in R-F scan and for $V_{\rm pol}<0$ in the F-R scan, while IH appears by inverting either the sign of poling voltage or the scan direction. Thus, we report both types of hysteresis, NH and IH, occurring in the same sample, under different poling conditions, but also in relation to the measurement scan directions.

\begin{figure}[t]
\centering
\includegraphics[width=8.5cm]{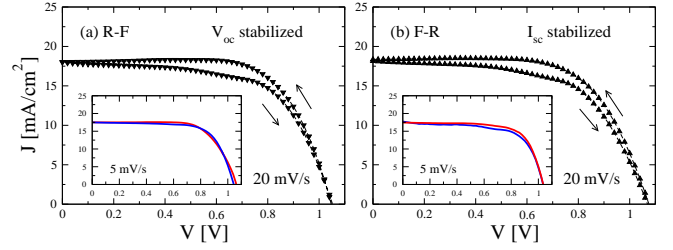}
\caption{J-V characteristics measured under stationary poling conditions, at 20 mV/s scan rate: (a) R-F scan, preceded by stabilization at $V_{\rm oc}$ and (b) F-R scan, after stabilization of $I_{\rm sc}$. The dynamic hysteresis is visibly reduced, compared to Fig.\ \ref{RF-FR-exp} for the same scan rate. The insets show almost hysteresis-free behavior for a lower scan rate of 5 mV/s.}
\label{minpol}
\end{figure}

Some degradation related effects as a consequence of repeated measurements may be observed, which bring up some variations compared to the ideal situation described in Fig.\ \ref{RF-FR}. Compared to the negative poling instances in Fig.\ \ref{RF-FR-exp}(b,d), which are performed first, the positively poled cells exhibit a drop in the open circuit bias as part of a temporary degradation process, from $V_{\rm oc} = 1.1$ V in the initial measurement, to $V'_{\rm oc} \simeq 0.98$ V. However, the bias scan range was set for all measurements to $V_{\rm min}=0$ and $V_{\rm max}=V_{\rm oc}$. Under these circumstances, as noted before, the degraded cells acquire additional poling in the voltage range $(V'_{\rm oc},V_{\rm oc})$ during both R-F or F-R scans. This is particularly noticeable for F-R scan ($P_0 = 10 P_\infty$), where current overshoot in the reverse scan is still visible.  
At the same time, one should note that reaching low negative values during individual reverse scans, performed at a rather small rate, can induce a negative poling for a subsequent scan. In this case, a second reverse scan will account for the negative polarization and the current shall be consequently reduced.

Although the data presented so far reflects rather extreme pre-poling conditions, these hysteretic behaviors can still be met to a finer degree in typical measurements. Of particular interest in establishing the maximum PCE are the J-V characteristics corresponding to stationary poling conditions. Figure\ \ref{minpol} presents J-V measurements for fresh samples stabilized at open circuit (R-F scan) and short circuit (F-R scan). Under these preconditioning steps, the bias scan rate of 20 mV/s induces a rather small hysteresis but, as long as the measurement time is still above the degradation characteristic time scale, smaller scan rates can further reduce the hysteresis loop. Here, the relatively large $I_{\rm sc}$ compared to the previous pre-poling measurements is indicative for the fact that the measurements were performed on electrically unstressed PSCs. Furthermore, a smaller rate of 5 mV/s reduces the hysteresis significantly, as shown in the inset of Fig.\ \ref{minpol}, rendering the samples almost hysteresis-free. 

\subsection{Hysteresis index}

To quantify the magnitude of the hysteretic effect several hysteresis indices (HIs) were proposed, based on the difference between the currents in the forward and reverse scans at given bias \cite{doi:10.1021/jz5011187,doi:10.1021/jz501392m,C7EE02048A}, on the ratio between the maximum power output of the two scans \cite{Calado2016,doi:10.1021/acs.chemmater.5b04019}, or on the difference between the integrated power output in reverse and forward scans \cite{doi:10.1021/acs.nanolett.7b01211}. We support the idea that the HI should be an integral measure reflecting the difference between the two scans, rather than a local property in the J-V curve. For the cases of {\it fully normal} and {\it fully inverted} hysteresis, i.e. when no crossing points between the forward and the reverse characteristics are present, we consider a definition in a slightly different form as in Ref.\ \cite{doi:10.1021/acs.nanolett.7b01211}:
\begin{equation}
{\mathcal H} = \frac{{\mathcal P}_{\rm oc\rightarrow sc} - {\mathcal P}_{\rm sc\rightarrow oc}}
                    {{\mathcal P}_{\rm oc\rightarrow sc} + {\mathcal P}_{\rm sc\rightarrow oc}}, 
\label{HI}
\end{equation}
where ${\mathcal P}_{\rm oc\rightarrow sc} = \int_{\rm sc}^{\rm oc} J_{\mbox{\scriptsize R}}(V) \Theta(J_{\mbox{\scriptsize R}}) dV$ and 
${\mathcal P}_{\rm sc\rightarrow oc} = \int_{\rm sc}^{\rm oc} J_{\mbox{\scriptsize F}}(V) \Theta(J_{\mbox{\scriptsize F}}) dV$ are the values of the integral power output in reverse and forward directions, respectively. $\Theta(J_{\mbox{\scriptsize F/R}})$ are Heaviside step functions. By construction, ${\mathcal H}$ is properly scaled, $-1\le {\mathcal H}\le 1$, positive values corresponding to NH, while negative ones to IH. Typical values are well below unity, while the limiting values $\pm 1$ can be obtained in extreme cases when either the current in the forward or in the reverse scan is negative, e.g. due to negative poling there is photovoltaic effect in only one scan direction. The choice for the denominator in Eq.\ (\ref{HI}) ensures a balanced evaluation of both NH and IH. Another advantage compared to HI definitions based on single working point, either using the reverse and forward currents or the PCEs, is related to a reliable HI value in the case of MH behavior.

\begin{table}
\caption{Hysteresis index for different poling regimes and measurement conditions, calculated for the experimental J-V characteristics.}
\scriptsize
\begin{tabular}{cccccc}
\hline\hline
Experimental  & Poling         & Scan        & Scan            & Hysteresis & Hysteresis \\
poling regime & bias           & direction   & rate            & type       & index \\
              & $V_{\rm pol}$ [V]  &             & $\alpha$ [mV/s] &            & $|{\mathcal H}|$ $[\%]$ \\
\hline
Strong poling            & 1.3 & R-F & 20  & NH & 12.6 \\ 
(20 mV/s)                & 1.3 & F-R & 20  & IH & 9.2 \\
$[$Fig.\ \ref{RF-FR-exp}$]$  &-1.5 & R-F & 20  & IH & 12.5 \\
                         &-1.5 & F-R & 20  & NH & 17.0 \\
\hline
Strong poling            & 1.3 & R-F &500  & NH & 2.2 \\ 
(500 mV/s)               & 1.3 & F-R &500  & IH & 1.5 \\
$[$Fig.\ \ref{RF-FR-exp}$]$  &-1.5 & R-F &500  & IH & 4.3 \\
                         &-1.5 & F-R &500  & NH & 4.1 \\
\hline
Moderate poling           & -2 & F-R & 20  & NH & 4.8 \\
$[$Fig.\ S2$]$            & -1 & F-R & 20  & NH & 4.3 \\
                          &  0 & F-R & 20  & MH & 3.5 \\
                          &0.5 & F-R & 20  & MH & 2.9 \\
                          &  1 & F-R & 20  & MH & 3.7 \\
                          &1.2 & F-R & 20  & MH & 6.6 \\
\hline
Stationary poling         & $V_{oc}$ & R-F & 20   & NH & 3.0 \\
$[$Fig.\ \ref{minpol}$]$      & 0        & F-R & 20   & NH & 3.4 \\
                          & $V_{oc}$ & R-F &  5   & MH & 1.3 \\
                          & 0        & F-R &  5   & MH & 1.7 \\
\hline
\hline

\end{tabular}
\label{tab1}
\end{table}

However, Eq.\ (\ref{HI}) can be consistently applied to R-F or F-R scans without crossing points, otherwise the resulting HI value would be misleadingly smaller due to partial canceling between regions of normal and inverted hysteresis in the J-V characteristics. We therefore propose a generalization of the HI definition to accommodate also the MH case with one or more crossing points $c_i$, by assigning one hysteresis index, ${\mathcal H}_i$ with $0\le i\le n$, for each bias interval, $(V_0=0,V_{c_1}),\ldots,(V_{i},V_{i+1}),\ldots,(V_n, V_{n+1})$, where $V_{n+1}=\max(V_{\rm oc}^R,V_{\rm oc}^F)$ is the largest value between the apparent open circuit voltage in the forward and reverse scans, $V_{\rm oc}^R$ and $V_{\rm oc}^F$, respectively. Hence, in general, we may characterize the dynamic J-V hysteresis by the $(n+1)$-tuple $({\mathcal H}_0,\ldots,{\mathcal H}_n)$ and the corresponding bias intervals, where:
\begin{equation}
{\mathcal H}_{i} = \frac{ {\mathcal P}_{ c_{i+1}\rightarrow {c_i}}
                    - {\mathcal P}_{ {c_i}\rightarrow {c_{i+1}}} }
                    {{\mathcal P}_{oc\rightarrow sc} + {\mathcal P}_{sc\rightarrow oc}}, 
\label{GHI}
\end{equation}
with ${\mathcal P}_{{c_i}\rightarrow {c_{i+1}}} = \int_{V_i}^{V_{i+1}} J_{\mbox{\scriptsize F}}(V) \Theta(J_{\mbox{\scriptsize F}}) dV$  and
${\mathcal P}_{{c_{i+1}}\rightarrow {c_{i}}} = \int_{V_i}^{V_{i+1}} J_{\mbox{\scriptsize R}}(V) \Theta(J_{\mbox{\scriptsize R}}) dV$.
Furthermore, one may define a single value associated with magnitude of MH as a sum of partial hysteresis indices, $|{\mathcal H}| = \sum_i |{\mathcal H}_i|$. This definition is equivalent to calculating the ratio between the total area enclosed by the forward and reverse scans, 
$\int_{V_0}^{V_{n+1}} | J_{\mbox{\scriptsize R}}(V) \Theta(J_{\mbox{\scriptsize R}}) - J_{\mbox{\scriptsize F}}(V) \Theta(J_{\mbox{\scriptsize F}}) | dV $, and the total integrated power output in the two scans, $\int_{V_0}^{V_{n+1}} ( J_{\mbox{\scriptsize R}}(V) \Theta(J_{\mbox{\scriptsize R}}) + J_{\mbox{\scriptsize F}}(V) \Theta(J_{\mbox{\scriptsize F}}) ) dV $. Note that for NH and IH Eq.\ (\ref{HI}) becomes a particular case of Eq.\ (\ref{GHI}).

At this point we provide in Table\ \ref{tab1} a condensed picture regarding the experimentally observed dynamic hysteresis effects, which are characterized by the hysteresis index defined in Eq.\ (\ref{GHI}). We distinguish between three different poling regimes, namely strong, moderate and stationary poling. 
Under strong poling conditions and a relatively slow scan rate ($\alpha=20$ mV/s) we obtain a quite large hysteresis with magnitudes in the range of 9-17\%, while for moderate poling $|{\mathcal H}|$ takes values of only 3-7\%, where MH is present with one crossing point. We indicate the crossing points $V_c$ and the MH indices $({\mathcal H}_0,{\mathcal H}_1)$ calculated for $V_{\rm pol}=0.5, 1.0, 1.2$V as shown in Fig.\ S2(d,e,f): $V_c=0.18$V, (-0.2\%, 2.7\%); $V_c=0.42$V, (-1.9\%, 1.7\%); $V_c=0.54$V, (-5.7\%, 0.9\%).
 Low $|{\mathcal H}|$ values of 2-4\% are also obtained for strong poling, but performing measurements at a fast scan rate ($\alpha=500$ mV/s). The stationary poling regime is closest to the stationary case, particularly for the lowest scan rate used, with a minimum $|{\mathcal H}|$ value of 1.3\% obtained for R-F scan with $\alpha=5$ mV/s, while the corresponding F-R scan indicates a slightly higher value of 1.7\% as a result of sample degradation. By establishing the stationary J-V characteristics, we find the maximum PCE of fresh samples at 12.6\%.

Our analysis also proves that although our samples present significant hysteresis under certain pre-poling conditions, the dynamic effects may be drastically reduced if proper pre-conditioning is performed. Hence, the reciprocal is also true and the lack of hysteresis in some reported conditions like slow scans of unpolarized samples or fast scan rates, does not at all exclude the possibility of obtaining large hysteresis. Therefore, without employing a proper MP, it remains unclear whether the samples exhibiting large hysteresis are not measured in optimal conditions and, conversely, whether apparently hysteresis-free PSCs would develop hysteresis under strong poling conditions.  
Presently, the MPs in focus are mostly concerned with a reliable determination of the maximum PCE. Yet it is equally important to accurately assess the hysteretic effects and this can be achieved by exploring a broad range of pre-poling conditions and, importantly, using correlated forward and reverse bias scans.

\subsection{MP guidelines}

Although techniques like MPPT \cite{C7TC03482B}, SCFV \cite{snaith,C7TA05609E} become more and more involved in the PCE evaluation for tackling the complex real-time response of the PSCs, dynamic J-V characteristics comprising of independent forward and reverse scans are still employed in the vast majority of studies. We provide in the following a set of guidelines to be considered for a proper characterization of the dynamic hysteresis, but also for an accurate PCE determination, which are accommodated within a measurement protocol with three distinct steps: $V_{\rm oc}$ stabilization, PSC poling under light soaking (optional) and actual measurement in the $(0,V_{\rm oc})$ bias range. These remarks, stemming from measurement conditions rather than structural variations of the PSCs, are well supported by both experiment and DEM simulations. 

\begin{itemize}
\item {\it Bias scan rate.} By increasing the bias scan rate the current is enhanced under positive PSC poling and reduced under negative poling \cite{doi:10.1021/acs.jpclett.5b00289,doi:10.1021/acs.jpcc.7b04248}. In order to obtain the static J-V characteristics, the optimal bias scan rate should be as low as possible, but without interfering with the degradation time scale. Very high scan rates can provide an apparent hysteresis-free behavior, which may not correspond to the stationary case and, for positive poling, may artificially enhance the PCE \cite{doi:10.1021/acs.jpclett.5b00289}. Since the hysteresis is maximized when $\tau\sim V_{\rm oc}/\alpha$ \cite{refId0} exploring only narrow ranges for the bias scan rates may render the hysteresis effect to be either increasing or decreasing with $\alpha$.  
\item {\it Bias scan range.} Typical measurements use a fixed bias scan range exceeding $(0,V_{oc})$. However, care must be taken as biases beyond $V_{\rm oc}$ may introduce strong positive poling. Similarly, applying low negative voltages may induce negative polarization. In this context, acquiring in advance the stabilized $V_{\rm oc}$ value is recommended.  
\item {\it Bias scan direction.} Usually the reverse scans overestimate the PCE, while the current is lower in the forward direction ({\it normal hysteresis}). However, depending also on the poling of the PSC, the opposite situation can occur: for strong negative poling one obtains {\it inverted hysteresis} \cite{doi:10.1021/acs.jpcc.7b04248}. At intermediate poling, a {\it mixed hysteresis} may be found, characterized by one or more crossing points between forward and reverse characteristics. Together with the magnitude and sign of the pre-poling voltage, the bias scan directions, R-F or F-R, set the hysteresis type. 
\item {\it Stabilized $V_{\rm oc}$.} Knowing the stabilized value of the open circuit bias is essential for setting the optimal bias scan range so that hysteretic effects are reduced, while using a low enough scan rate. In particular, $V_{\rm oc}$ may change over time, in successive measurements, as a consequence of PSC degradation, which may be temporary or permanent. 
\item {\it PSC poling.} The relation between the PSC poling on individual scans \cite{C4EE02465F} and on the type and magnitude of the dynamic hysteresis \cite{doi:10.1021/acs.jpcc.7b04248} is also well established, as well as other effects like the current overshoot in reverse due to positive initial polarization \cite{C4EE03664F,NEMNES2017197}. Yet, one should underline an often overlooked but rather important aspect: unintentional poling of the PSC may occur whenever the bias scan range is sensibly larger than $(0,V_{\rm oc})$; in this context, a changing $V_{\rm oc}$ brings additional complications.  
\item {\it J-V characteristics with stationary initial poling.} Following stabilization of $V_{\rm oc}$ a subsequent R-F scan provides a minimal NH at given scan rate. Alternatively, one can perform a F-R scan by first stabilizing the $I_{\rm sc}$ current. To obtain a J-V characteristics as close as possible to the stationary case, one should reduce the bias scan rate down to the limit set by the degradation time scale.
\item {\it Correlated forward and reverse bias scans.} Independent forward and reverse scans do not always provide an accurate description of the hysteretic effects, particularly when uncontrolled (unintentional) poling is involved. Consecutive R-F or F-R scans can instead provide a meaningful evaluation of the dynamic hysteresis, regarding both its type and magnitude.  
\item{\it Hysteresis index ${\mathcal H}$.} A wide variety of methods for calculating the hysteresis index have been proposed so far. We introduce here a properly scaled ${\mathcal H}$ index, as a global parameter for the entire working regime, with ${\mathcal H}>0$ for NH and ${\mathcal H}<0$ for IH. Furthermore, in the case of MH with $n$ crossing points, one may characterize the more complex behavior by $(n+1)$-tuple of hysteresis indices  ${\mathcal H}_i$. The overall magnitude can be assigned as the sum of $|{\mathcal H}_{i}|$'s calculated for the corresponding bias intervals.

\end{itemize}

\section{Conclusions}

In spite of the progress achieved for an accurate PCE determination, employing techniques like MPPT and SCFV, proper MPs for a consistent evaluation of the dynamic hysteresis need further attention. We investigated the dynamic effects induced by several measurement conditions, the numerical simulations in the framework of DEM being validated by experiment. We showed that both the scan direction in consecutive R-F and F-R scans and the sign of the bias pre-poling voltage set up the type of the hysteresis, normal or inverted. Furthermore, the hysteresis magnitude and short circuit current are influenced by the bias scan rate. In particular, at very large scan rates the J-V characteristics present small or no hysteresis, but the obtained results may be far away from the stationary case. The time spent between measurements susceptive of unintentional pre-poling may also change the type and magnitude of the hysteresis. In this context, a question which arises is in how far uncorrelated forward and reverse scans offer an accurate description of the hysteretic phenomena. 
Moreover, the usual procedure of setting a fixed bias scan range, without connecting it specifically to the open circuit bias, may yield misleading results: e.g. performing R-F scans for a group of devices with different $V_{\rm oc}$ or, similarly, for a single device with degradation affected $V_{\rm oc}$, one may obtain rather distinct hysteretic behavior, as a consequence of reaching different poling conditions at $V_{\rm oc}$. On the other hand, by stabilizing the PSC at $V_{\rm oc}$ or $V=0$, the two scans (R-F or F-R) offer a good approximation of the stationary case. Recurring features like the current overshoot in reverse and the crossing of the forward and reverse characteristics are analyzed in detail for experimentally relevant conditions and are consistently explained by DEM. We collect these observations in a set of guidelines for proper dynamic measurements. The investigation of dynamic effects is not only important for an accurate and efficient evaluation of the PCE, but also may give clues about the quality of the absorber layer and interfaces and, consequently, into the potential degradation of the solar cell. Therefore routine investigations beyond the stationary case are commendable as it may provide further insights into the performance and stability of PSCs.\\

{\bf Acknowledgements} \\

This work was supported by the National Ministry of Research and Innovation under the projects PN18-090205, PN18-110101 and by Romania-JINR cooperation project JINR Order 322/21.05.2018, no. 29.

\appendix
\section{Supplementary Material}

\begin{flushleft}
{\bf Contents:}\\ 
1. Fabrication and characterization methods.\\
2. Mixed hysteresis in simulated J-V characteristics.\\
3. Mixed hysteresis in experimental J-V characteristics.\\ 
\end{flushleft}


{\bf 1. Fabrication and characterization methods.}\\

 The PSCs were fabricated using a 100 nm TiO$_2$ compact layer deposited onto FTO coated commercial glass substrate (resistivity 7$\Omega$/sq, Solaronix TCO22-7) through spary-pyrolysis using a solution of titanium bis(acetylacetonate) solution (Aldrich) and nitrogen as a carrier gas, at 450 $^\circ$C. A mesoporous film was then obtained using a commercial TiO$_2$ paste (Solaronix Ti-Nanoxide N/SP) that was annealed at 500 $^\circ$C for 1h. The active layer consists of a mixed halide hybrid perovskite CH$_3$NH$_3$PbI$_{3-x}$Cl$_x$ that was prepared using a modified one-step method. Using a precursor solution (1107mg PbI$_2$ + 168mg PbCl$_2$ + 1800mg DMF + 210mg DMSO + 477mg CH$_3$NH$_3$I (Dysol)) a perovskite film was depositated by spin-coating at 2000 rpm for 25 s with the addition of 100 $\mu$l of diethyl ether at second 9 of the spin cycle onto the still spinning film, followed by a annealing step at 100 $^\circ$C for 3 minutes for final film crystallization. Also using spin-coating, a 200 nm thick spiro-OMeTAD layer was obtained at 1500 rpm for 30 s, the deposition using a solution that contained 80 mg spiro-OMeTAD (Borun Chemical), 28 $\mu$l 4-tert-butylpyridine and 18 $\mu$l of bis(trifluoromethane)sulfonimide lithium salt in acetonitrile solution (520 mg ml$^{-1}$) and the procedure being executed in a controlled atmosphere, at 24 $^\circ$C and under 10\% humidity. Gold was used as a counter electrode, a 100 nm thick film with an area of 0.09 cm$^2$ was deposited by RF magnetron sputtering technique. Further details regarding the fabrication and stability of the PSCs are described in Ref.\ \cite{doi:10.1021/acs.jpclett.6b02375}.

The J-V characteristics were measured using an Oriel VeraSol-2 Class AAA LED Solar Simulator having AM 1.5 filters and a Keithley 2400 Source Meter. The irradiation intensity was calibrated by a Newport standard silicon solar cell 91150 at 100 mW/cm$^2$. The 1 Sun illumination has been performed through a rectangular aperture of 3$\times$3 mm$^2$ size, of a geometry identical with that of Au counter electrodes.\\

{\bf 2. Mixed hysteresis in simulated J-V characteristics}\\

\begin{figure*}[h]
\centering
\includegraphics[width=14cm]{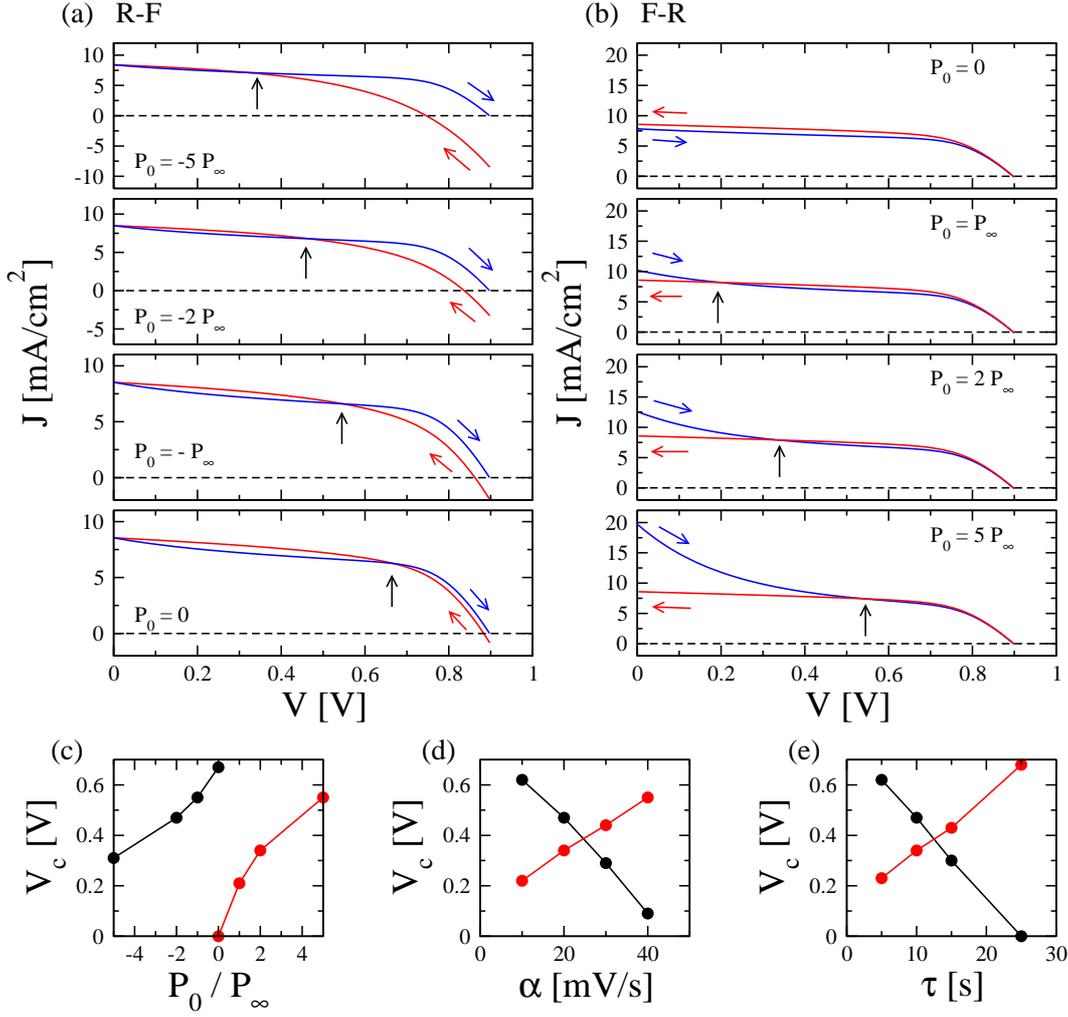}
\caption{Mixed hysteresis (MH) behavior under changing the poling conditions (a,b,c), bias scan rate (d) and relaxation time (e), for R-F and F-R scans. The positions of the crossing points ($V_c$) of the forward (blue) and reverse (red) J-V characteristics are marked by vertical arrows. The dependence of $V_c$ on $P_0$, $\alpha$ and $\tau$ is depicted for R-F (black) and F-R (red) in each case.}
\label{vc}
\end{figure*}

Figure\ \ref{vc}(a,b) shows simulated J-V characteristics using the dynamic electrical model (DEM) \cite{NEMNES2017197} obtained by varying the pre-poling, $P_0<0$ for the reverse-forward (R-F) scan and $P_0>0$ for the forward-reverse (F-R) scan, using a scan rate $\alpha = 20$ mV/s and considering a relaxation time $\tau = 10$ s. A mixed hysteresis (MH) is observed, characterized by a crossing point $V_c$. Increasing the $|P_0/P_\infty|$ ratio, the crossing point $V_c$ decreases towards short-circuit in the case of the R-F scan and oppositely for the F-R scan, the results being summarized in Fig.\ \ref{vc}(c). This is due to the fact that the reverse current in Fig.\ \ref{vc}(a) increases as the polarization is changed from $-5 P_\infty$ to 0, while in the F-R scan, the forward current is enhanced as $P_0$ is increased to $5 P_\infty$. Investigating the bias scan rate dependence of $V_c(\alpha)$, we notice a different behavior between the R-F and F-R scans. 
Increasing $\alpha$ in R-F scan with $P_0<0$, the initial negative polarization is present at lower biases, which keeps the reverse current below the forward one in a larger bias interval $(V_c,V_{\rm oc})$. Conversely, decreasing $\alpha$, $P_0$ decays in a smaller bias interval $(V_c,V_{\rm oc})$, and a normal hysteresis (NH) is restored for $V<V_c$. The opposite situation occurs for the F-R scan, where the increase of $\alpha$ drives $V_c$ towards $V_{\rm oc}$. 
The behavior of $V_c$ by changing $\alpha$, an extrinsic measurement protocol (MP) related parameter, is directly connected to the dependence of $V_c$ on $\tau$, an intrinsic, structure related parameter. For larger $\tau$ values and fixed $\alpha$, the decay of $P_0$ occurs over a larger bias interval, which, in the case of R-F scan decreases $V_c$, in contrast to the F-R case. Since $\tau$ and $\alpha$ are interrelated parameters, analyzing a broad sequence of scan rates can also provide a more accurate evaluation of the relaxation time scale.\\

\newpage

{\bf 3. Mixed hysteresis in experimental J-V characteristics}\\

\begin{figure*}[h]
\centering
\includegraphics[width=14cm]{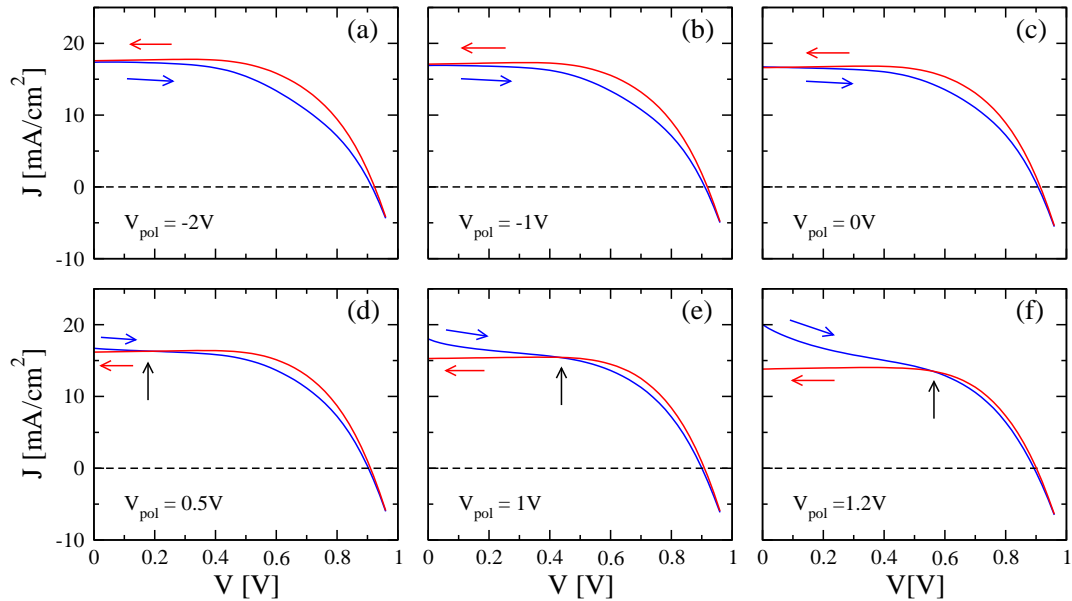}
\caption{Mixed hysteresis (MH) observed in experimental J-V characteristics in F-R scans, at 20 mV/s scan rate: tuning the hysteresis type by changing $V_{pol}$, from NH to MH and further towards complete IH. The vertical arrows mark the crossing points between forward and reverse characteristics.}
\label{cpexp}
\end{figure*}

We exemplify in Fig.\ \ref{cpexp} the tunability of the dynamic hysteresis induced by changing the pre-poling of the PSCs before the F-R scans. Starting with negative values for $V_{\rm pol}$, a relatively small NH is obtained. Typically, the negative poling condition is more difficult to achieve experimentally compared to the positive one. Increasing the pre-poling bias from -2V to 0V the NH is reduced, followed by MH for $V_{\rm pol}\ge0$. The $V_c$ voltage is shifted towards higher biases, as predicted by the DEM calculations in Fig.\ \ref{vc}(b). The MH can be ultimately turned into inverted hysteresis (IH) \cite{doi:10.1021/acs.jpcc.7b04248}, provided the positive poling is high enough. However, the relatively high current at biases larger than $V_{\rm oc}$ induce more and more PSC degradation, visible in the diminished $J_{\rm sc}$ on the reverse scan.\\

\balance

\newpage


\bibliography{manuscript_R2}

\end{document}